\documentclass[dvips]{article}
\usepackage{graphicx}
\usepackage{amssymb}
\usepackage{graphicx}
\usepackage{dcolumn}
\usepackage{amsmath}
\usepackage{lscape}
\usepackage{longtable}
\usepackage{subfigure}
\usepackage{color}

\begin{document}
\newcommand{\bd}{\begin{document}}
\newcommand{\ed}{\end{document}}
\newcommand{\bc}{\begin{center}}
\newcommand{\ec}{\end{center}}
\newcommand{\bfr}{\begin{flushright}}
\newcommand{\efr}{\end{flushright}}
\newcommand{\lt}{\left}
\newcommand{\rt}{\right}
\newcommand{\vs}{\vspace}
\newcommand{\hs}{\hspace}
\newcommand{\beq}{\begin{equation}}
\newcommand{\eeq}{\end{equation}}
\newcommand{\lb}{\linebreak}
\newcommand{\pb}{\pagebreak}
\newcommand{\mb}{\makebox}
\newcommand{\fb}{\framebox}
\newcommand{\mc}{\multicolumn}
\newcommand{\ben}{\begin{enumerate}}
\newcommand{\een}{\end{enumerate}}
\newcommand{\bit}{\begin{itemize}}
\newcommand{\eit}{\end{itemize}}
\newcommand{\ol}{\overline}
\newcommand{\un}{\underline}
\newcommand{\lefq}{\lefteqn}
\newcommand{\ba}{\begin{array}}
\newcommand{\ea}{\end{array}}
\newcommand{\beqa}{\begin{eqnarray}}
\newcommand{\eeqa}{\end{eqnarray}}
\newcommand{\beqas}{\begin{eqnarray*}}
\newcommand{\eeqas}{\end{eqnarray*}}
\newcommand{\bfg}{\begin{figure}}
\newcommand{\efg}{\end{figure}}
\newcommand{\bds}{\begin{displaymath}}
\newcommand{\eds}{\end{displaymath}}
\newcommand{\btb}{\begin{tabbing}}
\newcommand{\etb}{\end{tabbing}}

\title[\textbf{\huge  Supersymmetric Analysis of Spinning Cosmic String Spacetime Within External fields with Aharonov-Bohm interaction}

\bc
{\huge \"Ozlem Ye\c{s}ilta\c{s}$^{*}${\footnote {e-mail : yesiltas@gazi.edu.tr}
 and Bilgehan  B. \"{O}ner \\ Department of Physics, Faculty of Science, Gazi University,
06500 Ankara, Turkey\\
\vspace{.16cm}

}} \ec \vs{1cm}

\begin{abstract}
{The supersymmetric analysis of spinning cosmic string spacetime, involving an electron in magnetic fields, has been conducted. We examined the Dirac system within extended special functions known as exceptional orthogonal polynomials. Corresponding Dirac system is transformed to a relativistic system with a nonlinear isotonic oscillator. Furthermore, new potential models that extend the radial oscillator by adding rational terms are expressed in terms of the exceptional orthogonal Laguerre $X_{m}$ polynomial. The necessary analyses of the potential, energy levels, and probability density graphs are introduced for various cosmic string topological defects and Aharonov-Bohm interaction parameters.}
\end{abstract}

{ { \textbf{keywords}:   cosmic strings, supersymmetry, fermion, Aharonov-Bohm}}

\section{Introduction}
 Cosmic strings are hypothetical one-dimensional topological defects in the fabric of spacetime, predicted by some theories of the early universe. These strings are not "strings" in the traditional sense, but rather are incredibly thin regions of space where the field equations of a field theory have configurations that are drastically different from the vacuum state. If these strings could spin, they would have interesting gravitational properties due to their mass and angular momentum.

  Other examples of topological defects in physics can be listed as in low-temperature condensed matter physics, Bose-–Einstein condensates,  magnetic flux tubes in superconductors. The first introduce of hypothetical $1$-dimensional topological defects were done in the 1970s \cite{Kibble}. Then,  physical properties and the cosmological evolution of various defects are examined in \cite{vilenkin}. One of the importance in the topic of cosmic strings arises from the formation of large-scale structures such as galaxies. But,  when we question the existence of cosmic strings by searching for  an imprint in the cosmic microwave background (CMB), unfortunately it is revealed that there is not enough footprints of the cosmic strings in the CMB yet.
In case of interaction with the matter, they may deflect massless particles and emmit gravitational waves \cite{ry}.  The cosmic strings form closed loops  and gravitational waves emitted by loops. By the way, detectability of gravitational microlensing of distant quasars by cosmic strings is discussed in \cite{kon}. On the other hand, the possibility of generation of a primordial magnetic field by a network of charged-current carrying cosmic strings are explored \cite{m}, \cite{nayak}. It is argued that cosmic string may produce plasma vorticity and magnetic fields \cite{vilenkin}. Cosmic strings and magnetic field interactions are introduced in  the literature such as Aharonov-Bohm Interaction of Cosmic Strings with matter  \cite{Alford}, the quantization of a spin--0 matter field  in the background of a straight cosmic string \cite{sit}, the quantum dynamics of Dirac fermions in
presence of a cosmic string by introducing a conical topological defect in gapped graphene \cite{chak}, self-adjoint extensions \cite{silva}, vacuum currents induced by a magnetic flux \cite{bezerra}, relativistic quantum dynamics under external fields \cite{bezerra2}. Moreover, the relativistic quantum dynamics of Dirac oscillator on the curved spacetime of a rotating cosmic string is studied \cite{hassan}. In the following study \cite{silva2}, analytical approach to the spin-1/2 particle in the spinning cosmic string spacetime under magnetic fields and Aharonov-Bohm potential is performed to obtain radial equation solutions. A spinning cosmic string is the gravitational analog of the Aharonov-Bohm solenoid. The Aharonov-Bohm effect is a quantum mechanical phenomenon in which an electron is affected by an electromagnetic potential, despite being confined to regions where both the magnetic field and electric field are zero. In this scenario, an electron experiences a change in phase due to the electromagnetic potential, which can lead to observable interference effects \cite{R1}. This effect is a fundamental aspect of quantum mechanics and has important implications for the understanding of quantum fields and gauge theories. The Aharonov-Bohm potential specifically refers to the vector potential that influences the electron, even in regions where the magnetic field itself is not present. This concept emphasizes the non-locality in quantum mechanics and the significance of potentials in the quantum description of electromagnetism. It's a vivid demonstration that in quantum mechanics, it's not just the fields but the potentials that have physical significance \cite{R2}. When we consider an electron in an Aharonov-Bohm potential around a cosmic string, we're combining aspects of quantum mechanics with ideas from theoretical astrophysics. The cosmic string, due to its unique properties, would alter the spacetime around it, and this alteration would be reflected in the behavior of the electron. In this setup, the Aharonov-Bohm effect would manifest not just due to the electromagnetic potential but also due to the spacetime curvature induced by the cosmic string. The cosmic string's gravitational effects would influence the phase of the electron wave function \cite{R3}, leading to potentially observable quantum interference effects that are distinct from those seen in typical Aharonov-Bohm setups. When you combine these two concepts, a spinning cosmic string under the Aharonov-Bohm effect, it suggests a scenario where a cosmic string could influence charged particles not just through its gravitational field but also through electromagnetic effects in a quantum mechanical context \cite{R4}. This could lead to unique interactions where the properties of the cosmic string (like its spin and mass) could influence the phase of wave functions of particles encircling it, even if these particles are in a region where classical electromagnetic fields are zero.

 In line with these studies, the motivation of the recent paper is the supersymmetric approach to investigate how a spin$-1/2$ particle under magnetic fields experiences the Aharonov-Bohm potential  in  spinning cosmic string spacetime. Supersymmetric quantum mechanics (SUSY QM) is an area of theoretical physics that applies the principles of supersymmetry to quantum mechanics. Supersymmetry is a theoretical framework that proposes a type of symmetry between bosons (particles that follow Bose-Einstein statistics, like photons) and fermions (particles that follow Fermi-Dirac statistics, like electrons). In the context of quantum mechanics, this approach leads to some interesting and useful insights. SUSY QM was originally developed as a toy model to understand supersymmetry in the more complex setting of quantum field theory \cite{R5}, but it has since become an interesting area of study in its own right. One of its main attractions is that it provides a framework for analytically solving certain quantum mechanical problems that would be difficult to solve by other means. It also offers a deeper understanding of the algebraic structure underlying quantum mechanics. Recent research on supersymmetric quantum mechanics (SUSY QM) has delved into various sophisticated aspects of the theory, including the multiphoton algebra and coherent states in systems with infinite discrete spectra. One study explored the multiphoton algebras for one-dimensional Hamiltonians and their SUSY partners, generating multiphoton annihilation and creation operators. This research also developed the concept of multiphoton coherent states (MCS), which are eigenstates of the multiphoton annihilation operator, and examined the uncertainty relations for these states in the context of SUSY QM. This approach allows for a deeper understanding of the algebraic structure underlying supersymmetric quantum systems and provides a framework for analyzing the dynamical properties of such systems through the lens of multiphoton algebra and coherent states  \cite{R6}.

 Moreover, a new expanded potential model, along with solutions expressed in terms of exceptional orthogonal polynomials \cite{R7}, \cite{R8} is obtained for the model that involves spinning cosmic strings under external fields. Exceptional orthogonal polynomials (EOPs) are a class of polynomials that generalize classical orthogonal polynomials by allowing for the possibility of certain "gaps" in their degree sequence. Unlike the classical orthogonal polynomials, which form complete sets of solutions to second-order linear differential equations and have no gaps in their degrees (i.e., they include polynomials of every non-negative integer degree), EOPs can omit one or more degrees and still form complete, orthogonal sets with respect to some weight function. Unlike classical polynomials (Hermite, Laguerre, and Jacobi polynomials), EOPs have sequences of degrees that are not consecutive integers, introducing "gaps" in the sequence. They are often classified into families that generalize the classical families, such as the exceptional Hermite, Laguerre, and Jacobi polynomials. EOPs have attracted significant interest in mathematical physics, particularly in the context of exactly solvable models of quantum mechanics, where they are used to construct new solvable potentials and to study supersymmetric quantum mechanics \cite{R9}.

  This work is organized as follows: Section II details the algebra in curved spacetime, the separation of the Dirac equation and supersymmetry, and includes the primary solutions of the system and its rationally extended version. Section III discusses exceptional $X_\textbf{m}$ Laguerre polynomials, which are the solutions of the Dirac system. Finally, Section IV presents the conclusion of our results.

\section{Dirac Equation in the Gravitational Background With Rotational and Magnetic Field Effects}
The rotating cosmic sting spacetime metric is defined by \cite{Cunha}
\begin{equation}
ds^2=\sum_{ij} g_{ij} \, dx^i dx^j = (dt+a d\varphi)^{2}-dr^{2}-\alpha^{2}r^{2}d\varphi^{2}-dz^{2} \label{gammametric},
\end{equation}
where $\alpha$ is the parameter related to the the topological defect  $\alpha=1-4\mu$, $\mu$ is the linear mass density, $\mu \in (0,1]$, $a$ is the rotation parameter given in terms of angular momentum $J$,  $a=4J$. The interval for the parameters can be written as $-\infty <z <\infty$, $r\geq 0$, $0\leq \varphi \leq 2\pi$. The metric is proposed as \cite{bakke}
\begin{equation}\label{metric}
  ds^{2}=dt^{2}+2a dt d\varphi-(\alpha^{2}r^{2}-a^{2})d\varphi^{2}-dr^{2}-dz^{2},
\end{equation}
where the $\alpha^{2}r^{2}-a^{2} <0$ a condition related to closed timelike curves \cite{kp}, \cite{ring}, the spacetime admits closed timelike curves if $r < \frac{|a|}{\alpha}$. The dynamics equation for a spin-$1/2$ particle, considering its interaction with rotation and external magnetic fields, is given by
\begin{equation}\label{diraceq}
  i\gamma^{\mu}(x)(\nabla_{\mu}+i e A_{\mu}(x))\Psi(x)=M \Psi(x),
\end{equation}
where covariant derivaive is $\nabla_{\mu}=\partial_{\mu}+\Gamma_{\mu}$. Dirac matrices $\gamma^{\mu}(x)$ satisfy the expression
\begin{equation}\label{1}
  \gamma^{\mu}(x)=e_{a}^{\mu}(x)\gamma^{a},
\end{equation}
where $e^{\mu}_{a}$ are the tetrad fields, $\gamma^{a}$ are the flat matrices in Minkowski spacetime where
\begin{equation}\label{2}
 \gamma^{0}= \left(
    \begin{array}{cc}
      1 & 0 \\
      0 & -1 \\
    \end{array}
  \right),~~  \gamma^{i}= \left(
    \begin{array}{cc}
      0 & \sigma^{i} \\
      \sigma^{i} & 0 \\
    \end{array}
  \right).
\end{equation}
The anti-commutation of the Dirac matrices give metric tensor, i.e.
\begin{equation}\label{3}
  \{\gamma^{\mu}(x),\gamma^{\nu}(x)\}=2g^{\mu \nu }(x).
\end{equation}
In the covariant derivative definition, the spin affine connection reads as
\begin{equation}\label{4}
  \Gamma_{\mu}(x)=\frac{1}{4}\gamma^{a}\gamma^{b}e^{\mu}_{a}(x)(\partial_{\mu}e_{b\nu}(x)-\Gamma^{\sigma}_{\mu \nu}e_{b\sigma}(x)),
\end{equation}
where $\Gamma^{\sigma}_{\mu \nu}$ are the Christoffel symbols. It is understood that tetrads are coefficients of components of an orthogonal basis, and the metric tensor can be transformed from a coordinate basis to the tetrad basis.  Here, the tetrad basis is used as follows,
\begin{equation}\label{5}
  e^{a}_{\mu}(x)=\left(
                   \begin{array}{cccc}
                     1 & 0 & a & 0 \\
                     0 & \cos\varphi & -r\alpha \sin\varphi & 0 \\
                     0 & \sin\varphi & r\alpha \cos\varphi & 0 \\
                     0 & 0 & 0 & 1 \\
                   \end{array}
                 \right),~~e^{\mu}_{a}(x)=\left(
                   \begin{array}{cccc}
                     1 & \frac{a\sin\varphi}{r\alpha} & -\frac{a\cos\alpha}{r\alpha} & 0 \\
                     0 & \cos\varphi & \sin\varphi & 0 \\
                     0 & -\frac{\sin\varphi}{r\alpha} & \frac{ \cos\varphi}{r\alpha} & 0 \\
                     0 & 0 & 0 & 1 \\.
                   \end{array}
                 \right)
\end{equation}
Then, spinor affine connection reads as
\begin{equation}\label{6}
  \Gamma_{\mu}=(0,0,\frac{i}{2}(1-\alpha)\Sigma,0)
\end{equation}
where
\begin{equation}\label{7}
  \Sigma=\left(
           \begin{array}{cccc}
             1 & 0 & 0 & 0 \\
             0 & -1 & 0 & 0 \\
             0 & 0 & 1 & 0 \\
             0 & 0 & 0 & -1 \\
           \end{array}
         \right).
\end{equation}
Because (\ref{diraceq}) includes magnetic field interactions, vector potential can be expressed as
\begin{eqnarray}\label{07}
  \textbf{A} &=& [0,A_{\varphi},0] \\
  A_{\varphi}(r) &=& -\frac{1}{2}\alpha Br^{2}-\frac{\Phi}{e} \label{007}
\end{eqnarray}
where $\Phi$ is the magnetic flux through the solenoid. In the presence of $\textbf{A}$, as given in
(\ref{07}), the wavefunction of an electron is influenced by this vector potential, even the magnetic field
$\textbf{B} = \nabla \times \textbf{A} = B_{z} \hat{k} $ is  zero. $ A_{\varphi}(r)$, exhibiting cylindrical symmetry, would affect the particle's wavefunction in a manner similar to that in an electron-nucleus system. Therefore, the Aharonov-Bohm phenomenon is expected to occur around the cosmic string considered in this study. Next, the spinor of the fermion field is written as below
\begin{equation}\label{8}
  \Psi(t,r,\varphi)=\exp(-iEt)\left(
                                \begin{array}{c}
                                  \psi_{1}(r,\varphi) \\
                                  \psi_{2}(r,\varphi) \\
                                \end{array}
                              \right)
\end{equation}
where
\begin{eqnarray}
 \psi_{1}(r,\varphi)   &=& \left(
                             \begin{array}{c}
                               e^{im\varphi} \psi_{11}(r) \\
                               i e^{i(m+1)\varphi} \psi_{12}(r) \\
                             \end{array}
                           \right),~~~~ \psi_{2}(r,\varphi) = \left(
                             \begin{array}{c}
                               e^{im\varphi} \psi_{21}(r) \\
                               i e^{i(m+1)\varphi} \psi_{22}(r) \\
                             \end{array}
                           \right)
  \end{eqnarray}
Then, the Dirac system is separated as follows,
\begin{eqnarray}\label{D1}
  -\frac{1}{2M}u''_{11}+\left(\frac{\ell_{1}(\ell_{1}+1)}{2Mr^{2}}+V_{1}(r)\right)u_{11}(r) &=& \varepsilon^{2}u_{11}(r) \\
   -\frac{1}{2M}u''_{12}+\left(\frac{\ell_{2}(\ell_{2}+1)}{2Mr^{2}}+V_{2}(r)\right)u_{12}(r)  &=& \varepsilon^{2}u_{12}(r) \label{D2}
\end{eqnarray}
where $\varepsilon^{2}=E^{2}-M^{2}$ and,
\begin{eqnarray}\label{P1}
  V_1(r) &=& \frac{B^{2}e^{2}}{8M}r^{2}-\frac{Be}{2M}\left(\ell_{1}+\frac{3}{2}\right) \\
  V_2(r) &=&  \frac{B^{2}e^{2}}{8M}r^{2}-\frac{Be}{2M}\left(\ell_{2}+\frac{1}{2}\right). \label{P2}
\end{eqnarray}
Here, wavefunction transformations and angular momentum numbers are defined as follows,
\begin{eqnarray}
  u_{1i}(r) &=& \sqrt{r}\psi_{1i}(r),~~ i=1,2,\\
  \ell_{1} &=& \frac{1+2m-2\alpha+2a E-2\Phi}{2\alpha}, \\
  \ell_{2} &=& \frac{-1-2m-2\alpha-2a E+2\Phi}{2\alpha}.
 \end{eqnarray}
The system defined by equations (\ref{D1}) and (\ref{D2}), with the elements given in (\ref{P1}) and (\ref{P2}), is known as a three-dimensional oscillator. The solutions to this system are already known. So, the system, as presented in \cite{Q}, reads as follows
\begin{equation}\label{Q1}
  -y''(x)+V_{\ell}(x)y(x)=\epsilon y(x)
\end{equation}
\begin{equation}\label{Q2}
  V_{\ell}(x)=\frac{1}{4}\omega^{2}x^{2}+\frac{\ell(\ell+1)}{x^{2}},
\end{equation}
and the solutions are given by \cite{Q}

\begin{equation}\label{Q3}
 \epsilon^{\ell}_{n}=\omega\left(2n+\ell+\frac{3}{2}\right),~~~~n=0,1,2,...
\end{equation}
\begin{equation}\label{Q4}
  y_{n}(x)=\left( \frac{\omega}{2}\right)^{\frac{1}{2}(\ell+3/2)}\sqrt{\frac{2n!}{\Gamma(n+\ell+3/2)}}~ x^{\ell+1} e^{-\frac{1}{4}\omega x^{2}} L^{\ell+1/2}_{n}\left(\frac{1}{2}\omega x^{2}\right),
\end{equation}
where $L^{\ell+1/2}_{n}(\frac{1}{2}\omega x^{2})$ are the Laguerre polynomials. Now we can define (\ref{D1}) and (\ref{D2}) in terms of the new variable of $x$ which is $x=\sqrt{2M}r$ and obtain ,
\begin{eqnarray}\label{D11}
  -u''_{11}(x)+V_{eff}(x) u_{11}(x) &=& \epsilon^{2}u_{11}(x) \\
   -u''_{12}(x)+U_{eff}(x) u_{12}(x) &=& \epsilon^{2}u_{12}(x), \label{D12}
\end{eqnarray}
and
\begin{eqnarray}\label{Vef}
 V_{eff}(x) &=& \frac{\ell_{1}(\ell_{1}+1)}{x^{2}}+\frac{B^{2}e^{2}}{16M^{2}}x^{2}-\frac{Be}{2M}\left(\ell_{1}+\frac{3}{2}\right),  \\
  U_{eff}(x) &=& \frac{\ell_{2}(\ell_{2}+1)}{x^{2}}+\frac{B^{2}e^{2}}{16M^{2}}x^{2}-\frac{Be}{2M}\left(\ell_{2}+\frac{1}{2}\right) \label{Uef}
\end{eqnarray}
If we match (\ref{Q1}) and (\ref{D1}), we obtain the parameters $\{\omega, \ell \}$ in terms of our parameters as
\begin{eqnarray}
  \omega &=&  \frac{Be}{2M } \\
  \ell &=&   \ell_{1} ~~~~  or ~~~~ \ell=-1-\ell_{1}.
\end{eqnarray}
By the way, $\ell_{2}$ can be obtained from $\ell_{1}$ using the mirror symmetry of the parameters $E, m, \Phi$, i.e,
\begin{equation}\label{9}
  E\rightarrow -E, ~~ m\rightarrow -m,~~ \Phi \rightarrow -\Phi.
\end{equation}
Then, we can express $E_n$ and $u^{\ell_{1}}_{11,n}(r)$ given as below
\begin{eqnarray}\label{011}
   E^{\ell_{1}}_{n}  &=&  \frac{-aBe}{2M\alpha}\pm \frac{\sqrt{4a^{2}B^{2}e^{2}-8M\alpha \gamma}}{4M\alpha}, ~~\ell=-1-\ell_1 \label{0011}
\end{eqnarray}
where $\gamma=Be(1+2m-\alpha-2n\alpha-2\Phi)-2M^{3}\alpha$,

\begin{equation}\label{11}
  u^{\ell}_{11,n}(x)=\left(  \frac{Be}{2M}\right)^{\frac{1}{2}(\ell_{1}+3/2)}\sqrt{\frac{2n!}{\Gamma(n+\ell_{1}+3/2)}}~x^{\ell_{1}+1} e^{-\frac{1}{4} \frac{Be}{2M} x^{2}} L^{\ell_{1}+1/2}_{n}\left( \frac{Be}{4M } x^{2}\right).
\end{equation}
Here, (\ref{D1}) and (\ref{D2}) are known as shape invariant potentials in the literature \cite{suk}. In the standard SUSY QM, the superpotential is given in terms of the derivative of the groundstate solutions $W(r)=-\frac{u'_{11,0}(x)}{u_{11,0}(x)}$ and the partner potentials are expressed using the function $W(x)$,
\begin{eqnarray} \label{12}
V_{eff}(x) = V_1(x) &=& W^{2}(x)-\frac{dW}{dx}+E^{2} \\
  V_2(x) &=& W^{2}(x)+\frac{dW}{dx}+E^{2} \label{120}.
\end{eqnarray}
If we call the Hamiltonian operator for each system in (\ref{D1}) and (\ref{D2}) as
\begin{eqnarray}
  H_1 &=& \mathcal{A}^{\dag}\mathcal{A}=-\frac{d^{2}}{dx^{2}}+V_1(x)-E^{2} \\
  H_2 &=& \mathcal{A}\mathcal{A}^{\dag}=-\frac{d^{2}}{dx^{2}}+V_2(x)-E^{2}
\end{eqnarray}
and
\begin{equation}\label{13}
   \mathcal{A}=\frac{d}{dx}+W(x),~~\mathcal{A}^{\dag}=-\frac{d}{dx}+W(x)
\end{equation}
where   $E$ is the factorization energy, ~$\mathcal{A} u_{11,0}(x)=0$, $u_{11,0}(x)$ is the groundstate wavefunction of the corresponding $V_1(x)$ potential. Here, $V_2(x)$ shares the same energy spectrum with $V_1(x)$ except groundstate \cite{suk}. We can define now the superpotential as
\begin{equation}\label{14}
  W(x)=\frac{Be}{4M}x+\frac{\ell_{1}}{x},
\end{equation}
where we use the second solution to the angular momentum $\ell=-\ell_{1}-1$. Hence, we can continue with the extension of the radial oscillator classes of our system. Before, let's check the graph of potential functions and energy levels.

\begin{figure}
\begin{center}
\includegraphics[height=2in]{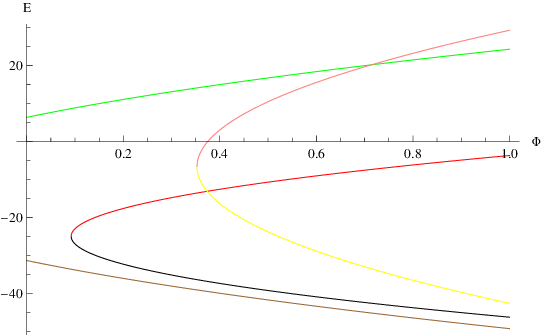}
\caption{Energy levels given in \ref{0011} with respect to parameter $\Phi$. $M=2, m=1, e=1, n=1$ for each curve while $(B,a,\alpha)=(100,0.1,0.1),(1,0.5,0.2), (8,0.1,0.45)$ for red/black, green/brown and  for pink and yellow curves. }
\end{center}
\end{figure}
In Figure $1$, it is observed that as the $\alpha$ topological deflection parameter with the angular momentum parameter $a$ increase, the energy levels become less negative.
\begin{figure}
\begin{center}
\includegraphics[height=2in]{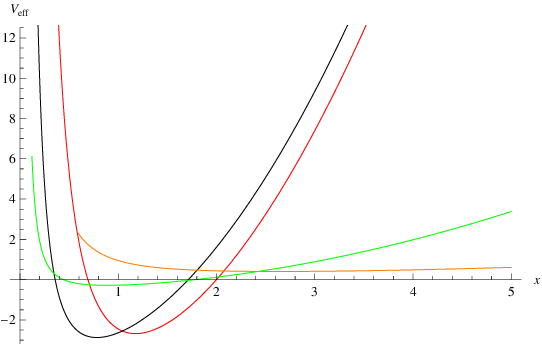}
\caption{Effective potential graph of (\ref{Vef})$M=2, m=1, e=1, B=10,\alpha=0.5$ for each curve while $(\Phi,a)=(0.4,1),( 5.0,2), (10,5),( 20,2)$ for red, black, green and orange correspondingly.  }
\end{center}
\end{figure}
Figure 2 shows that for smaller values of the magnetic flux parameter $\Phi$, the equilibrium point of the potential appears deeper. When $\Phi$ assumes dramatically greater values, the potential curve approaches zero more rapidly.
\begin{figure}
\begin{center}
\includegraphics[height=2in]{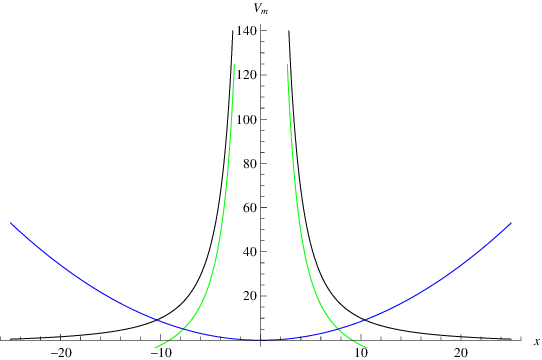}
\caption{Effective potential graph of (\ref{Uef})$M=2, m=1, e=1, n=1, \Phi=0.10, \alpha=0.1$ for each curve while ${(B,a)}={(0.1,1.2),( 0.5,10),( 140,50)}$ for  black, green and blue curves correspondingly.  }
\end{center}
\end{figure}

\begin{figure}
\begin{center}
\includegraphics[height=2in]{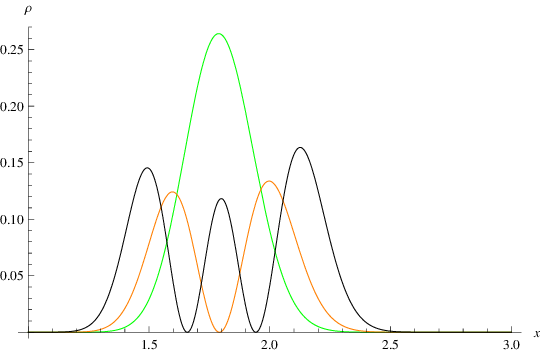}
\caption{Probability density $\rho(r)=|u_{11}|^{2}+|u_{12}|^{2}$ graphs where $u_{12}$ are the solutions of (\ref{D12}). $B=100, \alpha=0.1, \Phi=10, a=0.1$ and n=0,1,2 for green, orange and black curves correspondingly.  }
\end{center}
\end{figure}

\begin{figure}
\begin{center}
\includegraphics[height=2in]{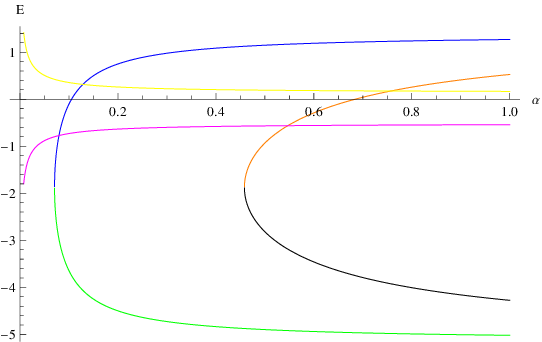}
\caption{Energy levels given in \ref{0011} with respect to parameter $\alpha$. $M=2, m=1, e=1, a=0.1$ for each curve while $(\Phi,B)=(0.5, 5),(1.3,10),(5.0,50)$ for orange/black,   blue/green and  for yellow/magenta curves respectively. }
\end{center}
\end{figure}

The topological deflection parameter is increased to $\alpha=0.9$ in Figure 3, which depicts the partner potential graph of (\ref{Uef}). As observed in the graph, as $B$ and $\alpha$ increase, the potential well becomes deeper. Figure $4$ shows the probability density graph is drawn for the solutions given in (\ref{11}).

In figure $5$, as the value of $\alpha$ increases, the energy becomes more stable.
\newpage
\subsection{Rationally extended radial oscillator model}
\emph{\textbf{ ansatz for the superpotential}}\\
Let us give an ansatz for the superpotential which may both  give $V_1(x)$ in (\ref{12}) under some  parameter restrictions and also produce the partner potential $V_{2,p}(x)$ which is different from the one given in (\ref{120}). Then, we have
\begin{equation}\label{15}
  W(x)=\frac{a_1}{x}+a_2 x-\frac{f'(x)}{f(x)}.
\end{equation}
This ansatz in (\ref{15}) is given in \cite{Q}, then, we can follow the straightforward algebra in \cite{Q}. Thus, one obtains $V_1(x)$ and $V_{2,p}(x)$ as follows
\begin{eqnarray}\label{16}
  V_1(x) &=& a_2(2a_1-1)+\frac{a_1(a_1+1)}{x^{2}}+a^{2}_{2}x^{2}+\frac{2a_1 f'(x)}{f(x)}+\frac{2a_2 xf'(x)}{f(x)}+\frac{2f'^{2}}{f^{2}}-\frac{f''}{f} \\
  V_{2,p}(x) &=&  a_2(2a_1+1)+\frac{a_1(a_1-1)}{x^{2}}+a^{2}_{2}x^{2}+\frac{2a_1 f'(x)}{f(x)}+\frac{2a_2 xf'(x)}{f(x)}+\frac{2f'^{2}}{f^{2}}+\frac{f''}{f}. \label{17}
\end{eqnarray}
Using (\ref{15}) in (\ref{12}) and (\ref{120}) and $f(x)=x^{2}+c$, one gets
\begin{equation}\label{17}
  V_1(x)=E^{2}+2a_1a_2-a_{2}+\frac{a_1(a_1+1)}{x^{2}}+a^{2}_{2}x^{2}-\frac{4a_{2}x^{2}+4a_1-2}{c+x^{2}},
\end{equation}
\begin{equation}\label{18}
  V_{2,p}(x)=E^{2}+2a_1a_2+a_{2}+\frac{a_1(a_1-1)}{x^{2}}+a^{2}_{2}x^{2}-\frac{4a_{2}x^{2}+4a_1+2}{c+x^{2}}+\frac{8x^{2}}{(c+x^{2})^{2}}.
\end{equation}
We can eliminate the rational and constant terms in (\ref{17}) using the parameters
\begin{equation}\label{19}
  c=\frac{2a_1-1}{2a_2},~~E^{2}=-a_2(2a_1-5).
\end{equation}
Then, $V_{2,p}(x)$ is obtained as
\begin{equation}\label{20}
  V_{2,p}(x)=V_{1}(x)+2W'(x)=a^{2}_{2}x^{2}+\frac{a_1(a_1-1)}{x^{2}}+\frac{4}{x^{2}+c}-\frac{8c}{(x^{2}+c)^{2}}+2a_2,
\end{equation}
where
\begin{equation}\label{21}
  a_1(a_1-1)=\ell_1(\ell_1+1),~~~~a^{2}_{2}=\frac{B^{2}e^{2}}{16M^{2}}.
\end{equation}
The solutions are given by $\{(a_1, a_2)=(-\ell_1,\frac{Be}{4M}), (-\ell_1, -\frac{Be}{4M}), (1+\ell_1, \frac{Be}{4M}), (1+\ell_1, -\frac{Be}{4M})\}$. Next, we can focus on the solutions of (\ref{20}). In the literature, (\ref{20}) is known as quantum isotonic nonlinear-oscillator potentials \cite{Hall}, \cite{sesma}. It is noted that  (\ref{20}) can be compared to Eq.$(28)$ in \cite{Hall}. If we apply the supersymmetric relationship of the wavefunctions
\begin{equation}\label{22}
  u_{12}(x)=C_{2} \mathcal{A}^{\dag}u_{11}=C_{2}\left(-\frac{d}{dx}+\frac{\ell_{1}+1}{x}+\frac{Be}{4M}x-\frac{2x}{x^{2}+c}\right)u_{11}(x)
\end{equation}
where $u_{11}(x)$ is given by (\ref{11}). Using  the differential identity of Laguerre polynomials given below
\begin{equation}\label{23}
  \frac{d}{dz}L^{\alpha}_{n}(z)=-L^{\alpha+1}_{n-1}(z),
\end{equation}
and
\begin{equation}\label{24}
  L^{\alpha}_{n}(z)=L^{\alpha}_{n-1}+L^{\alpha-1}_{n}(z),
\end{equation}
we obtain
\begin{equation}\label{25}
  u_{12}(x)=C_2 \frac{x^{\ell_{1}+2}e^{-\frac{Bex^{2}}{8M}}}{x^{2}+c^{2}}\left((2\ell_{1}+2n+5)L^{\ell_{1}+1/2}_{n}(x')
  -2(n+1)L^{\ell_{1}+1/2}_{n+1}(x')+\frac{Bec^{2}}{4M}L^{\ell_{1}+3/2}_{n}(x')\right)
\end{equation}
where $x'=\frac{Be}{4M}x^{2}$.
\\
 Figure $6$ is drawn for (\ref{20}) which also  shows that a potential well can be obtained for the smaller values of $\Phi$. Upon examining Figure $7$ it is seen that the rationally extended potential in equation (\ref{44}) differs from those shown previously. For smaller $\Phi$ values, such as $\Phi=0.1$ there is singularity at the points $|x|=7.797$.  Figure $8$ presents the probability densities for the solutions given in equations (\ref{Uef}) and (\ref{25}) while Figure $8$  shows the probability densities for the solutions given in equations $V_{_{\textbf{m}}}(x)$ and (\ref{25}).
\begin{figure}
\begin{center}
\includegraphics[height=2in]{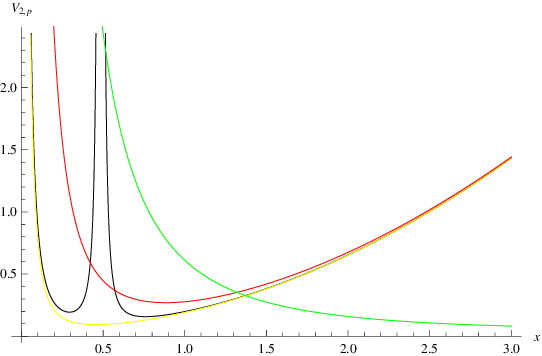}
\caption{Effective partner potential $V_{2,p}(x)$ graphs. $\Phi=0.5,1.5,0.1,50$ corresponding to black, yellow,red, green curves.}
\end{center}
\end{figure}
\newpage

\section{Exceptional Orthogonal Polynomials }
\subsection{$X_\textbf{m}$ exceptional Laguerre polynomials}
$X_1$-Jacobi and $X_1$- Laguerre are known as exceptional orthogonal polynomials which exist as two infinite sequences of polynomial eigenfunctions of a Sturm–Liouville problem. Unlike the classical orthogonal polynomial systems, their sequences start with a polynomial of degree one. These polynomials $R(g(x))$ satisfy a second order differential equation which is given by
\begin{equation}\label{30}
  R''(g(x))+F(g)R'(g(x))+G(g)R(g(x))=0.
\end{equation}
Here, $F(g)$ and $G(g)$ functions are given as
\begin{eqnarray}
  F(g) &=&\frac{1}{g}\left ((\alpha'+1-g)-2g\frac{L^{\alpha'}_{\textbf{m}-1}(-g)}{L^{\alpha'-1}_{\textbf{m}}}(-g)\right) \\
  G(g) &=& \frac{1}{g}\left(n-2\alpha'\frac{L^{\alpha'}_{\textbf{m}-1}(-g)}{L^{\alpha'-1}_{\textbf{m}}(-g)}\right),~~\textbf{m}\geq 0,~~n\geq \textbf{m},~~\alpha'>0.
\end{eqnarray}
The norm integral is given by
\begin{equation}\label{31}
  \int^{\infty}_{0} (L^{\alpha'}_{n,\textbf{m}}(g))^{2}W^{\alpha'}_{\textbf{m}}(g)dg=\frac{(\alpha'+n)\Gamma(\alpha'+n+\textbf{m})}{(n-\textbf{m})!},
\end{equation}
where
\begin{equation}\label{32}
  W^{\alpha'}_{n}(g)=\frac{g^{\alpha'}e^{-g}}{(L^{\alpha'-1}_{\textbf{m}})^{2}}
\end{equation}
and $X_m$ Laguerre polynomials can be written as
\begin{equation}\label{33}
  L^{\alpha'}_{n,\textbf{m}}=L^{\alpha'}_{\textbf{m}}(-g)L^{\alpha'-1}_{n-\textbf{m}}+L^{\alpha'-1}_{\textbf{m}}(-g)L^{\alpha'}_{n-\textbf{m}-1},~~n\geq \textbf{m}.
\end{equation}
$X_m$-exceptional orthogonal Laguerre polynomials  satisfies the differential equation below
\begin{equation}\label{34}
 \frac{d^{2} L^{\alpha'}_{n,\textbf{m}}}{dx^{2}}+\frac{1}{g}\left((\alpha'+1-g)-2g\frac{L^{\alpha'}_{\textbf{m}-1}(-g)}{L^{\alpha'-1}_{\textbf{m}}(-g)}\right)\frac{dL^{\alpha'}_{n,\textbf{m}}(g)}{dx}+
 \frac{1}{g}\left(n-2\alpha'\frac{L^{\alpha'}_{\textbf{m}-1}(-g)}{L^{\alpha'-1}_{\textbf{m}}}\right)L^{\alpha'}_{n,\textbf{m}}(g)=0.
\end{equation}
\subsection{initial equation}
Let us mention (\ref{30}) again. We need to find an equation of our system which fits (\ref{30}). When we consider the system (\ref{D11}), there should be a transformation in order to make (\ref{D11})-(\ref{D12}) comparable with (\ref{30}). So we apply $u_{12}(x)=x \chi(x)$, (\ref{D11}) turns into
\begin{equation}\label{35}
  \chi^{''}(x)+\frac{2}{x}\chi^{'}(x)+(\varepsilon^{2}-V_{eff}(x))\chi(x)=0.
\end{equation}
Then, we proceed with the point canonical transformation $\chi(x)=f(x)R(g(x))$   which is substituted into equation (\ref{35}) \cite{sud}, yielding
\begin{equation}\label{36}
  f(x)=x^{-1}(g^{'}(x))^{-1/2}\exp\left(\frac{1}{2}\int Q(x)dx\right),
\end{equation}
\begin{equation}\label{37}
  \varepsilon^{2}-V_{eff}(x)=\frac{1}{2}\frac{g'''}{g'}-\frac{3}{4}\frac{g''^{2}}{g'^{2}}+g'^{2}\left(R(g)-Q'(g)-\frac{Q^{2}}{4}\right).
\end{equation}
Now $Q(g)$ and $R(g)$ correspond to
\begin{eqnarray}\label{38}
  Q(g) &=& \frac{1}{g}\left((\alpha'+1-g)-2g\frac{L^{\alpha'}_{\textbf{m}-1}(-g)}{L^{\alpha'-1}_{\textbf{m}}(-g)}\right), \\
  R(g) &=&  \frac{1}{g}\left(n-2\alpha'\frac{L^{\alpha'}_{\textbf{m}-1}(-g)}{L^{\alpha'-1}_{\textbf{m}}}\right).    \label{39}
\end{eqnarray}
Hence, one can use $Q(g)$ and $R(g)$  in (\ref{37}), and get
\begin{equation}\label{39}
    \begin{split}
         \varepsilon^{2}-V_{eff, \textbf{m}}(x)&= \frac{1}{2}\frac{g'''}{g'}-\frac{3}{4}\frac{g''^{2}}{g'^{2}}+ \\
        & g'^{2}\left(-\frac{1}{4}+\frac{2n+\alpha'+1}{2g}-\frac{\alpha'^{2}-1}{4g^{2}}+\frac{L^{\alpha'+1}_{\textbf{m}-2}(-g)}{L^{\alpha'-1}_{\textbf{m}}}
   -\frac{\alpha'+g-1}{g}\frac{L^{\alpha'}_{\textbf{m}-1}(-g)}{L^{\alpha'-1}_{\textbf{m}}}-
   2\left(\frac{L^{\alpha'}_{\textbf{m}-1}(-g)}{L^{\alpha'-1}_{\textbf{m}}(-g)}\right)^{2}\right).
    \end{split}
\end{equation}
On the left handside of (\ref{39}), the constant $\varepsilon^{2}$ may be obtained using
\begin{equation}\label{40}
   \frac{g'^{2}}{g}=C,
\end{equation}
where $C$ is a constant. Since a term that is constant and aligns with $\varepsilon^{2}$, is required, the proposed form for $g(x)$  is presented in equation (\ref{40}). Then, $g(x)=\frac{1}{4}Cx^{2}$ can be obtained and $\varepsilon^{2}$ may be taken as
\begin{equation}\label{41}
  \varepsilon^{2}=n C,~~n=0,1,2,...
\end{equation}
using $n\rightarrow n+\textbf{m}$. Consequently, $V_{\textbf{m}}(x)$  is expressed as follows
\begin{equation}\label{42}
  V_{\textbf{m}}(x)=\frac{C^{2}}{16}x^{2}+\frac{\alpha'^{2}-1/4}{x^{2}}-
  \frac{C^{2}x^{2}}{4}\frac{L^{\alpha'+1}_{\textbf{m}-2}(-g)}{L^{\alpha'-1}_{\textbf{m}}(-g)}+
  C\left(\alpha'+\frac{Cx^{2}}{4}-1\right)\frac{L^{\alpha'}_{\textbf{m}-1}(-g)}{L^{\alpha'-1}_{\textbf{m}}(-g)}+\frac{C^{2}x^{2}}{2}
  \left(\frac{L^{\alpha'}_{\textbf{m-1}}(-g)}{L^{\alpha'-1}_{\textbf{m}}(-g)}\right)^{2}-\frac{C}{2}(2\textbf{m}+\alpha'+1).
\end{equation}
And, in our system, the constants $C$ and $\alpha'$ stand for
\begin{equation}\label{43}
  C=\frac{Be}{M}, ~~~~\alpha'=\frac{1}{2}(1+2\ell_{2}).
\end{equation}
Rewriting $V_{\textbf{m}}(x)$ in terms of the parameters of our system yields,

\begin{equation}\label{44}
    \begin{split}
       V_{\textbf{m}}(x) &= \frac{B^{2}e^{2}}{16}x^{2}+\frac{\ell_{2}(\ell_{2}+1)}{x^{2}}-
  \frac{B^{2}e^{2}}{4M^{2}}x^{2}\frac{L^{\ell_{2}+3/2}_{\textbf{m}-2}(-g)}{L^{\ell_{2}-1/2}_{\textbf{m}}(-g)}+
  \frac{Be}{M}\left(\ell_{2}-\frac{1}{2}+\frac{Be}{4M}x^{2}\right)\frac{L^{\ell_{2}+1/2}_{\textbf{m}-1}}{L^{\ell_{2}-1/2}_{\textbf{m}}}+ \\
        &  \frac{B^{2}e^{2}}{2M^{2}}\left(\frac{L^{\ell_{2}+1/2}_{\textbf{m}-1}}{L^{\ell_{2}-1/2}_{\textbf{m}}}\right)-\frac{Be}{2M}\left(2\textbf{m}+
        \ell_{2}+\frac{3}{2}\right).
    \end{split}
\end{equation}
By substituting  $Q(g)$ and $g(x)$ into (\ref{36}) and employing this in the expression $u_{12}(x)=x \chi(x)$, one obtains:
\begin{equation}\label{45}
  u_{12}(x)=N_{n,\textbf{m}} \frac{x^{\ell_{2}+1}e^{-\frac{Bex^{2}}{16M}}}
  {L^{\ell_{2}-1/2}_{\textbf{m}}(-\frac{Bex^{2}}{8M})}L^{\ell_{2}+1/2}_{n+\textbf{m},\textbf{m}}\left(\frac{Bex^{2}}{8M}\right).
\end{equation}

\begin{figure}
\begin{center}
\includegraphics[height=2in]{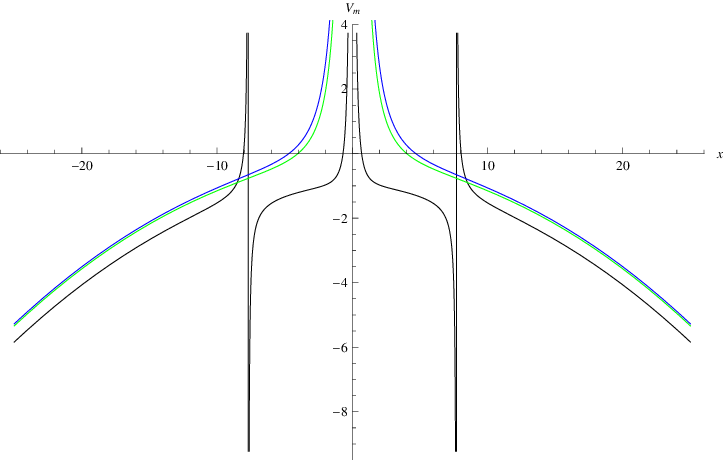}
\caption{Effective partner potential $V_{\textbf{m}}(x)$ in (\ref{44}) graphs. $\Phi=0.1,0.5,0.95$ corresponding to black, green, blue curves.}
\end{center}
\end{figure}

\begin{figure}
\begin{center}
\includegraphics[height=2in]{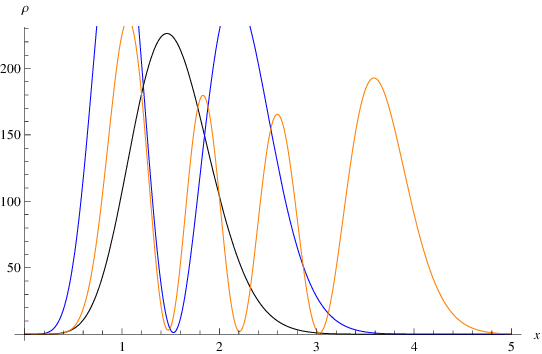}
\caption{Probability density $\rho(r)=|u_{11}|^{2}+|u_{12}|^{2}$ graphs where $u_{12}$ is given by (\ref{25}). $a=0.1, B = 10, M = 2, m = 1, \alpha = 0.2, \Phi = 0.9, e =
 1 $ and $n=0, 1, 3$ corresponding to black, blue, orange curves.}
\end{center}
\end{figure}

\begin{figure}
\begin{center}
\includegraphics[height=2in]{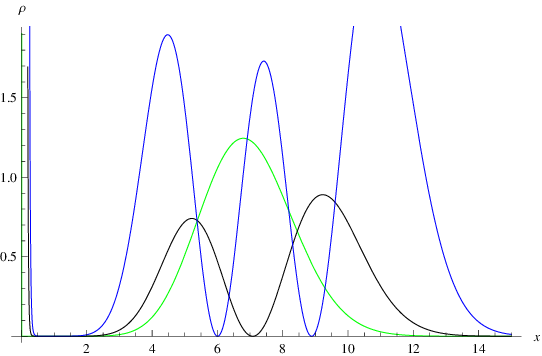}
\caption{Probability density $\rho(r)=|u_{11}|^{2}+|u_{12}|^{2}$ graphs where $u_{12}$ is given by (\ref{45}). $a=0.1, B = 10, M = 2, m = 1,\textbf{m}=1, \alpha = 0.2, \Phi = 0.5, e =
 1 $ and $n=0, 1, 2$ corresponding to green, black and blue curves correspondingly.}
\end{center}
\end{figure}

\begin{figure}
\begin{center}
\includegraphics[height=2in]{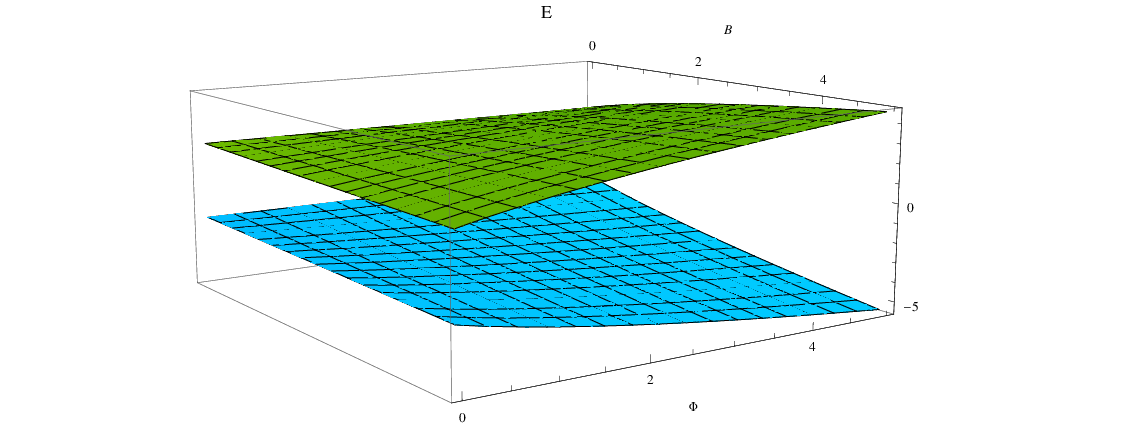}
\caption{Energy levels given in (\ref{0011}) as a function of $B$ and $\Phi$. $a=0.2, B = 10, M = 2, m = 1, \alpha = 0.6, n=2, e =
 1 $.}
\end{center}
\end{figure}
 Finally, the probability density for the solutions in (\ref{45}) and (\ref{11}), and energy as a function of magnetic field strength and flux of the solenoid, are given in Figures $9$ and $10$, respectively.

\newpage

\section{Conclusions}

This study demonstrates how the system of a spinning cosmic string spacetime, interacting with external fields via Aharonov-Bohm interaction, is adapted to rationally extended potential systems. Within these systems, the solutions of the partner Dirac Hamiltonian are found using SUSY QM and are expressed in terms of  $X_{m}$ exceptional orthogonal Laguerre polynomials. The newly generated family of rationally expanded potentials has been identified as belonging to the class of nonlinear isotonic potentials. According to the analytical solutions, a sudden drop in magnetic field strength can lead to shifts and transitions in energy levels  in Figure $1$ , which also agree with the results in \cite{silva2}. It is also observed that when $\alpha < 0.45$  and at smaller values of the magnetic field strength and flux, there are fewer positive energy levels compared to the energy of the antiparticles in Figure $5$. Moreover, the probability density graphs, which are obtained only for the smaller magnetic field strength and magnetic flux, are presented in Figures 8 and 9, in contrast to the probability density graph depicting greater magnetic field strength and magnetic flux in Figure 4. When we   examine the change in energy levels with respect to $B$ and $\Phi$ in the Figure $10$, we find that the antiparticle's eigenvalues are more numerous than those of the particle.
By expressing wavefunction solutions in terms of exceptional $X_m$ Laguerre polynomials, we have both the advantage of deriving new exactly solvable potentials in (\ref{44}) in terms of the string and magnetic field parameters, and it may also allow us to study systems with spectra that have "gaps" or missing levels, which can be related to the study of supersymmetric quantum mechanics.

\section{  Data Availability Statement }

Data sets generated during and/or analyzed during the current study are available from the corresponding author upon reasonable request. The data is housed within a secure repository to ensure compliance with ethical guidelines and privacy laws applicable to the study. Interested researchers may contact the corresponding author, providing a detailed explanation of their request and intended use of the data. Approval for data access will be subject to an assessment of the request's alignment with the conditions under which the data was collected and any ethical or legal restrictions.

\end{document}